\documentclass[11pt,letterpaper]{article}
\usepackage{soul}
\usepackage{jheppub}
\usepackage[utf8]{inputenc}
\usepackage[T1]{fontenc}
\usepackage{amsfonts}
\usepackage{tikz}
\usepackage{amsmath}
\usepackage{graphicx,subfigure}
\usepackage{tensor} 
\usepackage{mathtools} 
\usepackage{amssymb}
\usepackage{enumerate}
\usepackage{euscript}
\usepackage{mathrsfs}
\usepackage{hyperref}
\usepackage{indentfirst}
\usepackage{color}
\usepackage{float}
\usepackage{braket}

\newcommand{\nn}{\nonumber}

\newcommand{\ds}{{\rm dS}}
\newcommand{\lds}{\ell_{\rm dS}}
\newcommand{\eff}{{\rm eff}}

\title{ \boldmath Toward a Unified de Sitter Holography: A Composite $T\bar{T}$ and $T\bar{T}+\Lambda_2$ Flow
}

\author[]{Jing-Cheng Chang$^a$$^b$,}
\author[]{Yang He$^a$$^b$,}
\author[]{Yu-Xiao Liu$^a$$^b$\footnote{Corresponding author},}
\author[]{Yuan Sun$^c$\footnote{Corresponding author}}

\emailAdd{120220908561@lzu.edu.cn, heyang2023@lzu.edu.cn,  liuyx@lzu.edu.cn, sunyuan@csu.edu.cn}

\affiliation[a]{Lanzhou Center for Theoretical Physics, Key Laboratory of Theoretical Physics of Gansu Province,\\ Key Laboratory of Quantum Theory and Applications of MoE, \\Gansu Provincial Research Center for Basic Disciplines of Quantum Physics, \\ Lanzhou University, Lanzhou 730000, China
	\vspace{0.1cm}}

\affiliation[b]{Institute of Theoretical Physics $\&$ Research Center of Gravitation, \\School of Physical Science and Technology, Lanzhou University, Lanzhou 730000, China
	\vspace{0.1cm}}

\affiliation[c]{Institute of Quantum Physics, School of Physics, Central South University, Changsha 418003, China}

\abstract{In de Sitter (dS) holography, both the dS/CFT correspondence and the dS static patch holography have been extensively studied. In these two holographic frameworks, the dual field theories are defined on spacelike and timelike boundaries, respectively, where the inward motion of the holographic boundary into the bulk corresponds to the $T\bar{T}$ and $T\bar{T}+\Lambda_2$ deformations in the respective dual field theories. In this work, we develop a unified framework for these two dS holographic models by introducing a composite flow that incorporates both $T\bar{T}$ and $T\bar{T}+\Lambda_2$ deformations. We propose that this composite flow corresponds to the inward motion of a spacelike boundary from the asymptotic infinity of dS spacetime, traversing the cosmological horizon and approaching the worldline of a static observer. This proposal is supported by the computation of the quasi-local energy and the holographic entanglement entropy within the dS static spacetime and its extended geometry.}


\begin{document}
	\maketitle
	\flushbottom
	
	
	\section{Introduction}
	The holographic duality principle~\cite{tHooft:1993dmi,Susskind:1994vu} offers a fundamental framework for studying quantum gravity. The duality between $(d+1)$-dimensional Anti-de Sitter (AdS) spacetime and $d$-dimensional conformal field theory (CFT) was well established~\cite{Maldacena:1997re,Gubser:1998bc,Witten:1998qj}. Recent developments have further extended this framework to include the $T\bar{T}$-deformed CFT~\cite{Smirnov:2016lqw,Cavaglia:2016oda}, whose holographic dual was described by an AdS spacetime with a finite radial cutoff~\cite{McGough:2016lol,Kraus:2018xrn}\footnote{Further interesting studies related to the $T\bar{T}$ deformation can be found in the reviews~\cite{He:2025ppz,Jiang:2019epa}.
	}. In contrast, the holographic description of de Sitter (dS) spacetime remains an open and intriguing area of research~\cite{Witten:2001kn,Chakraborty:2023yed,Dey:2024zjx,Yadav:2024ray,Goodhew:2024eup,Thavanesan:2025kyc,Chakravarty:2025sbg}.

	In dS holography, two principal holographic frameworks have been proposed: the dS/CFT correspondence~\cite{Strominger:2001pn,Maldacena:2002vr,Anninos:2011ui} and the dS static patch holography~\cite{Anninos:2011af,Anninos:2017hhn,Susskind:2021omt,Susskind:2021esx,Susskind:2021dfc,Shaghoulian:2022fop,Shaghoulian:2021cef,Franken:2023pni,Franken:2023jas,Anninos:2024wpy,Ruan:2025uhl}. In the dS/CFT correspondence, the dual field theory is proposed to reside at the future spacelike asymptotic boundary, as inferred from the analysis of asymptotic symmetries~\cite{Strominger:2001pn}. In contrast, based on Bousso's covariant entropy bound~\cite{Bousso:1999xy}, Refs.~\cite{Susskind:2021omt,Susskind:2021esx,Susskind:2021dfc,Shaghoulian:2022fop,Shaghoulian:2021cef,Anninos:2024wpy,Ruan:2025uhl} argued that the dual field theory should be located on the stretched horizon within the framework of dS static holography.
	Depending on the position of the holographic boundary, holographic models in dS spacetime can be broadly classified into two categories. The first class corresponds to scenarios where the dual field theory resides on a spacelike boundary, as in the dS/CFT correspondence and  Cauchy slice holography~\cite{Araujo-Regado:2022gvw,Araujo-Regado:2022jpj,Araujo-Regado:2025elv}, where the dual theory is nonunitary~\cite{Hikida:2022ltr,Araujo-Regado:2022gvw,Doi:2022iyj,Araujo-Regado:2022jpj}. The second class introduces a timelike boundary in dS spacetime and places the dual field theory on this boundary, including worldline holography~\cite{Anninos:2011af,Anninos:2017hhn}, dS static patch holography~\cite{Susskind:2021omt}, the dS/dS correspondence~\cite{Alishahiha:2004md,Alishahiha:2005dj,Dong:2018cuv,Geng:2019bnn,Grieninger:2019zts}, and the half-dS holographic model~\cite{Kawamoto:2023nki}. In these cases, the dual field theory typically exhibits nonlocal features~\cite{Miyaji:2015yva,Geng:2019ruz,Geng:2020kxh,Kawamoto:2023nki,Chang:2024voo}. Moreover, a shift of the holographic boundary inward within the bulk induces different deformations on the field theory side. Specifically, the inward motion of a spacelike boundary corresponds to the $T\bar{T}$ deformation~\cite{Chen:2023eic,Chang:2024voo}, whereas the inward motion of a timelike boundary corresponds to the $T\bar{T}+\Lambda_2$ deformation~\cite{Gorbenko:2018oov,Lewkowycz:2019xse,Shyam:2021ciy,Coleman:2021nor,Torroba:2022jrk,Silverstein:2022dfj,Batra:2024kjl,Aguilar-Gutierrez:2024nst}.

	Following the proposal of the $T\bar T$ deformation in the context of the AdS/CFT correspondence, deformations of dS holographic models within the dS/dS correspondence have also been investigated. It was proposed that, in dS holography, shifting the timelike boundary inward into the bulk is dual to the $T\bar T+\Lambda_2$ deformation on the field theory side, with the $\Lambda_2$ term originating from the positive cosmological constant of dS spacetime~\cite{Gorbenko:2018oov}. This proposal is supported, in particular, by the trace flow equation
		\begin{equation}
			\tilde{T}^{i}_i=-\frac{\ell_{\text{(A)dS}}}{16\pi G}\tilde{\mathcal{R}}^{(2)} -4\pi G\ell_{\text{(A)dS}}\left(\tilde{T}^{ij}\tilde{T}_{ij}- (\tilde{T}^{i}_i)^2\right) - \frac{\eta-1}{8\pi G\ell_{\text{(A)dS}}}\,,
		\end{equation}
		where $\tilde{T}_{ij}$ and $\tilde{\mathcal{R}}^{(2)}$ denote the boundary Brown--York tensor and the intrinsic curvature of the timelike boundary, respectively, while $\ell_{\text{(A)dS}}$ represents the curvature radius of (A)dS spacetime. The parameter $\eta$ is related to the cosmological constant through $\Lambda_2=-\eta/\ell_{\text{(A)dS}}^2$. For AdS spacetime one has $\eta=1$ and $\Lambda_2=-1/(\ell_{\text{AdS}})^2$, whereas for dS spacetime $\eta=-1$ and $\Lambda_2=1/\ell_{\text{dS}}^2$. This implies that, upon identifying the corresponding bulk and boundary quantities, the trace flow equation on the dual field theory side of dS spacetime takes the form
		\begin{equation}
			T^{i}_i = -\frac{c}{24\pi}\mathcal{R}^{(2)} - 4\pi\lambda\,\langle T\bar{T}\rangle+\frac{2}{\pi\lambda}\,.
		\end{equation}
		Here $T_{ij}$ and $\mathcal{R}^{(2)}$ denote the stress tensor of the boundary field theory and the intrinsic curvature of the manifold on which the field theory is defined, respectively, and they differ from the corresponding bulk quantities $\tilde{T}_{ij}$ and $\tilde{\mathcal{R}}^{(2)}$ by Weyl factors, since they do not depend on the radial coordinate. Compared to the standard $T\bar{T}$ deformation, this expression contains an additional term associated with the positive cosmological constant. As a direct consequence of the trace flow equation, the partition function of the dual field theory satisfies the flow equation
		\begin{equation}
			\frac{\partial}{\partial\lambda}\log Z=-2\pi\int d^2x\sqrt{g}\langle T\bar{T}\rangle+\frac{1}{\pi\lambda^2}\int d^2x\sqrt{g}\,.
		\end{equation}
		This proposal was further supported by consistency checks based on holographic entanglement entropy~\cite{Lewkowycz:2019xse}. Moreover, within the framework of dS static patch holography, the entropy of the stretched horizon and its logarithmic corrections were computed by employing the $T\bar T+\Lambda_2$ deformation~\cite{Shyam:2021ciy}.
		
		Subsequently, deformations of the dS/CFT correspondence were explored in Ref.~\cite{Chen:2023eic}. It was proposed that the $T\bar T$ deformation of the dual field theory is holographically dual to moving the spacelike boundary at future infinity inward into the bulk. In contrast to the AdS/CFT case, both the deformation parameter and the central charge in the dS/CFT setup are purely imaginary. It was further shown in Ref.~\cite{Chang:2024voo} that the trace flow equation in dS/CFT can be obtained via analytic continuation of the corresponding trace flow equation in AdS/CFT through a double Wick rotation, which naturally renders both quantities imaginary. To test the proposed holographic duality of the deformed field theory, Ref.~\cite{Chen:2023eic} employed the sphere partition function to compute exactly the entanglement entropy of a meridian on the boundary sphere. These results were subsequently generalized in Ref.~\cite{Chang:2024voo} to a variety of dS backgrounds, including dS global spacetime and the dS Poincaré patch, where consistency between holographic entanglement entropy and geodesic lengths was examined.

	Composite flows formed by combining the $T\bar T$ flow with the $T\bar T+\Lambda_2$ flow have found wide-ranging applications. In Ref.~\cite{Gorbenko:2018oov}, the construction begins with a $T\bar T$ flow; as the deformation parameter increases, the timelike dS$_2$ boundary, on which the $T\bar T$-deformed CFT resides, moves from the AdS boundary toward the deep infrared. In this limit, by exploiting the indistinguishability of the AdS/dS and dS/dS geometries in the deep infrared regime, the $\Lambda_2$ term is introduced, and the $T\bar T+\Lambda_2$ flow emerges as the deformation parameter decreases. In essence, such a composite flow provides a mechanism for reconstructing dS spacetime. Moreover, Ref.~\cite{Coleman:2021nor} similarly exploited the near-horizon indistinguishability between the black hole horizon and the cosmological horizon to perform a microscopic counting of states in dS spacetime, thereby providing a microscopic description of the cosmological horizon entropy. In summary, the extended flow proposed above connects the two deformations listed in the second column of Table~\ref{summary}.

	In contrast to the construction joining AdS and dS throats discussed in~\cite{Gorbenko:2018oov}, the $T\bar{T}+\Lambda_2$ deformation can also arise in AdS spacetime with a spacelike boundary, as demonstrated by Ahmad, Almheiri, and Lin~\cite{AliAhmad:2025kki}. They employed a combined sequence of $T\bar{T}$ and $T\bar{T}+\Lambda_2$ deformations to probe the interior regions of a rotating BTZ black hole. In this setup, the $\Lambda_2$ term naturally emerges when the $T\bar{T}$ deformation pushes the spacetime boundary into the region between the outer and inner horizons, in which region, the boundary turns to spacelike while it is timelike outside the outer horizon. As the spacelike boundary is further pushed inward by the $T\bar{T}+\Lambda_2$ deformation, the $\Lambda_2$ term eventually vanishes once the boundary crosses the inner horizon and becomes timelike. Finally, the timelike boundary is then driven inward by the $T\bar{T}$ deformation all the way toward the black hole singularity. The two deformations that appear in the above analysis correspond to the two entries listed in the second row of Table~\ref{summary}\footnote{The $T\bar T+\Lambda_2$ deformation denotes a class of standard $T\bar T$ deformation plus a constant term, and is not restricted to dS spacetime.}.

	\begin{table}[h]
		\centering
		\begin{tabular}{|c|c|c|}
			\hline
			& Timelike boundary & Spacelike boundary  \\ 
			\hline
			AdS bulk & $T\bar{T}$ & $T\bar{T}+\Lambda_2$   \\ 
			\hline
			dS bulk& $T\bar{T}+\Lambda_2$ & $T\bar{T}$ \\ 
			\hline
		\end{tabular}
		\caption{Field-theoretic deformations corresponding to the inward motion of different boundaries in AdS and dS spacetimes.}
		\label{summary}
	\end{table}
	
	As discussed above, the two classes of models in dS holography exhibit pronounced differences in the fundamental properties of their dual field theories. Holography with a spacelike boundary is typically associated with a non-unitary dual field theory, whereas holography with a timelike boundary often gives rise to a dual theory that is intrinsically nonlocal. This inherent tension between unitarity and locality constitutes one of the central features that distinguish dS holography from the AdS/CFT correspondence, and it highlights the fundamental challenges in constructing a consistent holographic dual description of dS spacetime. Motivated by Ref.~\cite{AliAhmad:2025kki}, we propose a consistent dS holographic model by combining the $T\bar{T}$ and $T\bar{T}+\Lambda_2$ flows in dS spacetime, as summarized in the third row of Table~\ref{summary}. This construction provides a unified framework that simultaneously incorporates both classes of dS holographic models.
	\begin{figure}[H]
		\centering
		\subfigure[The Penrose diagram of dS spacetime.]{
			\includegraphics[scale=0.8]{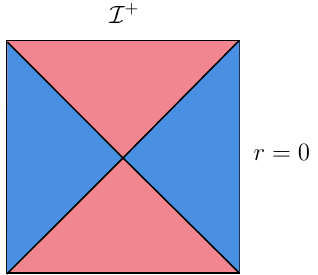}
			\label{dsstaticcoordinate}}
		\hspace{1in}
		\subfigure[$T\bar{T}$ flow and $T\bar{T}+\Lambda_2$ flow.]{
			\includegraphics[scale=0.8]{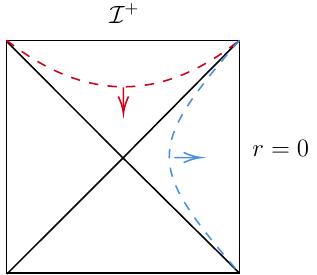}
			\label{dsstaticflow}
		}
		\caption{\ref{dsstaticcoordinate} The blue region denotes the static patch ($r<\lds$), the red region corresponds to the extended spacetime ($r>\lds$), and the interface represents the cosmological horizon at $r=\lds$. \ref{dsstaticflow} The red dashed line indicates the spacelike boundary, while the blue dashed line indicates the timelike boundary. Arrows represent the motion of these boundaries, which is opposite to the direction of their outward-pointing unit normal vectors.}
		\label{The Penrose diagram of dS static spacetime}
	\end{figure}
	Our construction is carried out in the static patch of dS spacetime and its extension. The static coordinate for dS spacetime is given by
	\begin{equation}
		\label{ds static coordinate}
		ds^2=-\left(1-\frac{r^2}{\lds^2}\right)dt^2+\left(1-\frac{r^2}{\lds^2}\right)^{-1}dr^2+r^2d\phi^2,
	\end{equation}
	where $r \in (0,\lds)$ corresponds to the static region, whereas $r \in (\lds,\infty)$ corresponds to its extended spacetime. The Penrose diagram of the dS static spacetime is shown in Fig.~\ref{The Penrose diagram of dS static spacetime}. In Fig.~\ref{dsstaticcoordinate}, $\mathcal{I}^{+}$ denotes the future infinity $r = \infty$, and $r=0$ represents the south pole, i.e., the position of a static observer. It is evident that outside the cosmological horizon ($r>\lds$), the coordinate $r$ behaves as timelike, while it remains spacelike inside the horizon. When the holographic principle is applied to the dS static spacetime, the holographic model outside the cosmological horizon corresponds to the first type of dS holographic model, whereas the model inside the horizon corresponds to the second type. In both cases, the dual field theory is defined on a codimension-one hypersurface at constant $r=r_c$. As illustrated in Fig.~\ref{dsstaticflow}, when the boundary starts from spacelike infinity, it is driven inward by the $T\bar{T}$ deformation. Once the boundary crosses the cosmological horizon and becomes timelike, the deformation transitions to the $T\bar{T}+\Lambda_2$ one, which continues to push the boundary toward the static observer. This process offers a unified description of the two classes of holographic models in dS spacetime.

	This paper is organized as follows. In Section~\ref{section2:flowequation}, we employ the Hamiltonian constraints to derive the trace flow equations for dS spacetimes with spacelike and timelike boundaries, and derive the associated deformation flow equations on the field theory side. In Section~\ref{Energy spectrum}, we compute the energy spectrum of the two deformation flows and match them at the critical deformation parameter through appropriate boundary conditions. In Section~\ref{Bulk analysic}, we provide a holographic interpretation of the composite flow formed by the $T\bar{T}$ and $T\bar{T}+\Lambda_2$ deformations, and test our proposal by evaluating the quasi-local energy. In Section~\ref{Holographic Entanglement Entropy}, we further examine our proposal through holographic entanglement entropy. Section~\ref{Conclusion and Discussion} presents our conclusions and discusses possible directions for future research.

	\section{$T\bar{T}$ and $T\bar{T}+\Lambda_2$ deformations in dS spacetime}
	\label{section2:flowequation}

	Given the elusive nature of the dual field theory in dS holography, we assume the validity of the holographic principle in dS spacetime and derive the corresponding deformation flow equations from the bulk perspective. The standard dictionary  of holographic duality identifies~\cite{Witten:1998qj,Gubser:1998bc} 
	\begin{equation}
		\label{holographic dictionary}
		Z_{\text{CFT}}[\gamma_{ab}] = \lim_{r_c \to \infty} Z_{\text{gravity}}[h^0_{ab} = r_c^2 \gamma_{ab}],
	\end{equation}
	where $Z_{\text{gravity}}$ denotes the gravitational partition function. In the framework of quantum gravity, the partition function is required to satisfy the Wheeler--DeWitt equation, which in this context takes the form $H Z_{\text{gravity}} = 0$~\cite{DeWitt:1967yk}. The Hamiltonian $H$ is obtained through a $2+1$ decomposition of spacetime
	\begin{equation}
		H = \frac{16\pi G}{\sqrt{|h|}} \left[\Pi^{ab}\Pi_{ab} - ({\Pi^a}_a)^2\right] + n^a n_a \frac{\sqrt{|h|}}{16\pi G} \left(\tilde{\mathcal{R}}^{(2)} - 2\Lambda_2\right) = 0,
	\end{equation}
	where $\Lambda_2$ is the cosmological constant, given by $\Lambda_2 = \frac{1}{\lds^2}$, and $n^a$ is the unit outward-pointing normal vector of the boundary at $r = r_c$, i.e., the direction towards increasing $r$. $\tilde{\mathcal{R}}^{(2)}$ denotes the intrinsic curvature of the boundary. And $\Pi^{ab}$ denotes the canonical momentum, which is defined as
	\begin{equation}
		\Pi^{ab} = n^a n_a\frac{\sqrt{|h|}}{16\pi G}(K^{ab} - K h^{ab}).
	\end{equation}
	Substituting the explicit form of the canonical momentum yields the Hamiltonian constraint equation
	\begin{equation}
		\label{WdWequation}
		-n^a n_a \tilde{\mathcal{R}}^{(2)} + K^2 - K_{ab} K^{ab} + 2\Lambda_2 n^a n_a = 0.
	\end{equation}
	This equation is consistent with the radial component of the Einstein field equations. By rewriting the Hamiltonian constraint equation in terms of the Brown--York tensor, one can straightforwardly obtain the trace flow equation in the bulk~\cite{Kraus:2018xrn}. Finally, invoking the holographic dictionary~\eqref{holographic dictionary} enables us to derive the corresponding deformation equation on the field theory side~\cite{McGough:2016lol,Hartman:2018tkw,Taylor:2018xcy}. We next discuss the deformation equations for the two cases, $n^a n_a = -1$ and $n^a n_a = 1$.

	\subsection{$T\bar{T}$ deformation with spacelike boundary}
	
	The analysis begins within the standard framework of the dS/CFT correspondence, wherein the cosmological wavefunction is dual to the partition function of the boundary field theory~\cite{Maldacena:2002vr},
	\begin{equation}
		Z_{\text{CFT}}\equiv \Psi_{\text{dS}}.
	\end{equation}
	Here the left-hand side denotes the partition function of the dual field theory, whereas the right-hand side represents the cosmological wavefunction of dS spacetime, which can be represented as a gravitational path integral,
	\begin{equation}
		\Psi_{\text{dS}}=Z_{\text{dS}}=\int \mathcal{D}\psi\, e^{-iS_{\text{dS}}},
	\end{equation}
	where $S_{\text{dS}}$ denotes the on-shell action including the Gibbons--Hawking--York boundary term\footnote{According to Stokes’ theorem, the counterterm for a spacelike boundary acquires an overall minus sign.}:
	\begin{equation}
			S_{\rm dS} =-\frac{1}{16\pi G}\int_{\mathcal{M}} d^3x\, \sqrt{-g} \left(\mathcal{R}^{(3)} - 2\Lambda\right) 
			+\frac{1}{8\pi G}\int_{\partial\mathcal{M}} d^2x\, \sqrt{h} \left(K - \frac{1}{\lds}\right).
	\end{equation}
	In the dS/CFT correspondence, since the CFT resides at future spacelike infinity with a Euclidean boundary metric, the Brown--York tensor of the Euclidean boundary~\cite{Brown:1992br,Balasubramanian:1999re} is defined as
	\begin{equation}
		\label{eq:BY define}
		\begin{aligned}
			\widetilde{T}_{(E)\,ab} &= \frac{2}{\sqrt{h}}\frac{\delta}{\delta h^{ab}}(-\ln Z_{\text{dS}}) \\[3pt]
			&= \frac{2i}{\sqrt{h}}\frac{\delta}{\delta h^{ab}}(S_{\text{dS}}^{\text{on-shell}}) \\[3pt]
			&= \frac{i}{8\pi G}\left(K_{ab} - K h_{ab} + \frac{h_{ab}}{\lds}\right),
		\end{aligned}
	\end{equation}
	where the subscript $(E)$ denotes that the stress tensor is defined in Euclidean signature, while $(L)$ for Lorentzian signature in the next subsection. For a spacelike boundary with $n^a n_a = -1$, the Hamiltonian constraint~\eqref{WdWequation} reduces to the standard $2+1$ decomposition form,
	\begin{equation}
		\widetilde{\mathcal{R}}^{(2)} + K^2 - K_{ab}K^{ab} - 2\Lambda_2 = 0.
	\end{equation}
	Rewriting the extrinsic curvature in terms of the Brown--York tensor yields the bulk trace flow equation,
	\begin{equation}
		\label{eq:traceflowequ-bulk}
		\widetilde{T}_{(E)}= \frac{i\lds}{16\pi G}\widetilde{\mathcal{R}}^{(2)} + i4\pi G\lds\left(\widetilde{T}_{(E)}^{ab}\widetilde{T}_{(E)\,ab} - (\widetilde{T}_{(E)})^2\right).
	\end{equation}
	Comparing this with the trace flow equation in the dual field theory \footnote{
		Through out this work, the tilde quantity   $\widetilde{T}_{ab}$ and untilde one $T_{ab}$ denote the stress tensor computed from bulk and boundary field theory, respectively.},
	\begin{equation}
		\label{eq:traceflowequ-CFT}
			T_{(E)}= -\frac{c}{24\pi}\mathcal{R}^{(2)} - 4\pi\lambda\,\langle T\bar{T}\rangle, \quad
			T\bar{T} \equiv \frac{1}{8}\left(T_{(E)}^{ab}T_{(E)\,ab} - T_{(E)}^2\right),
	\end{equation}
	one can then establish the holographic dictionary that relates bulk and boundary quantities~\cite{Chen:2023eic},
	\begin{equation}
		\label{dsconstrant}
		c = -i\frac{3\lds}{2G} = -i c_{\text{dS}}, \qquad 
		\lambda = -i\frac{8G\lds}{r_c^2} = -i\lambda_{\text{dS}}.
	\end{equation}

	It should be noted that the Brown--York tensor and the stress tensor of the dual CFT differ by a conformal factor depending on the radial coordinate $r$. Near spacelike infinity, the static coordinates of dS spacetime can be written in the Fefferman--Graham form,
	\begin{equation}
		ds^2 = -\frac{\lds^2}{r^2}dr^2 + h_{ab}dx^a dx^b = -\frac{\lds^2}{r^2}dr^2 + r^2\gamma_{ab}dx^a dx^b,
	\end{equation}
	where $\gamma_{ab}$ denotes the physical boundary metric. In this framework, the relations between bulk and boundary quantities are given by~\cite{Hartman:2018tkw,AliAhmad:2025kki}
	\begin{equation}
		\label{dictionary}
			\widetilde{\mathcal{R}}^{(2)} = \frac{1}{r^2}\mathcal{R}^{(2)},\;
			\widetilde{T}_{(E)\,ab} = T_{(E)\,ab}, \;
			\widetilde{T}_{(E)}^{ab} = \frac{1}{r^4}T_{(E)}^{ab}, \; 
			\widetilde{T}_{(E)} = \frac{1}{r^2}T_{(E)}.
	\end{equation}
	Substituting the holographic dictionary~\eqref{dictionary} into the bulk trace flow equation~\eqref{eq:traceflowequ-bulk} precisely reproduces the corresponding trace flow equation on the field theory side~\eqref{eq:traceflowequ-CFT}. For a flat boundary, we focus on the trace contribution induced by the deformation operator,
	\begin{equation}
		\label{eq:trace relation}
		T_{(E)} = -\frac{\pi\lambda}{2}\left(T_{(E)}^{ab}T_{(E)\,ab} - T_{(E)}^2\right).
	\end{equation}
	To derive the flow equation for the partition function, we consider a fluctuation of the deformation parameter, $\lambda + \delta\lambda$. By employing the diffeomorphism invariance of the boundary induced metric, one obtains the infinitesimal coordinate transformation
	\begin{equation}
		x^a \to x^a \left(1+\frac{\delta\lambda}{2\lambda}\right),
	\end{equation}
	which leads to the corresponding fluctuation of the physical boundary metric,
	\begin{equation}
		\delta\gamma_{ab} = -\frac{\delta\lambda}{\lambda}\gamma_{ab}, \qquad 
		\delta\gamma^{ab} = \frac{\delta\lambda}{\lambda}\gamma^{ab}.
	\end{equation}
	Using the definition of the stress tensor in~\eqref{eq:BY define}, we obtain
	\begin{equation}
		\label{variation of Z}
			\delta(-\ln Z_{\text{QFT}})=\frac{1}{2}\int d^2x \sqrt{\gamma}\,T_{(E)\,ab}\delta\gamma^{ab}
			=\frac{1}{2\lambda}\int d^2x \sqrt{\gamma}\,T_{(E)}\,\delta\lambda.
	\end{equation}
	Therefore, the flow equation for the partition function takes the form
	\begin{equation}
		\label{partition flow equation}
			\partial_{\lambda}\ln Z_{\text{QFT}}=-\frac{1}{2\lambda}\int d^2x \sqrt{\gamma}\,T_{(E)}=\frac{\pi}{4}\int d^2x\sqrt{\gamma}\left(T_{(E)}^{ab}T_{(E)\,ab} - T_{(E)}^2\right).
	\end{equation}
	Since the deformation parameter in the flow equation~(\ref{partition flow equation}) is complex, by substituting the saddle-point approximation of the partition function $Z_{\text{QFT}} = e^{-iS_{\text{QFT}}}$ together with the duality relation~(\ref{dsconstrant}), we then obtain the flow equation for the action with respect to the real deformation parameter $\lambda_{\ds}$ as
	\begin{equation}
		\label{the first step}
		\partial_{\lambda_\ds}S_{\text{QFT}} = \frac{\pi}{4} \int d^2x\,\sqrt{\gamma}\,\left(T_{(E)}^{ab}T_{(E)\,ab} - T_{(E)}^2\right).
	\end{equation}

	\subsection{$T\bar{T}+\Lambda_2$ deformation with timelike boundary}
	
	We now turn to the case where a timelike boundary is introduced in the dS spacetime. The on-shell action of dS spacetime with such a timelike boundary can be written as
	\begin{equation}
			S_{\rm dS} = -\frac{1}{16\pi G}\int_{\mathcal{M}} d^3x \sqrt{-g} \left(\mathcal{R}^{(3)} - \frac{2}{\lds^2}\right)
			- \frac{1}{8\pi G}\int_{\partial\mathcal{M}} d^2x \sqrt{-h} \left(K - \frac{1}{\lds}\right).
	\end{equation}
	Since the dual field theory resides on the timelike boundary, the Brown--York tensor is defined with Lorentzian signature\footnote{Our convention is $Z_{\text{dS}} = \int \mathcal{D}\psi\, e^{-iS_{\text{dS}}}$.} as
	\begin{equation}
		\label{eq:BY define timelike}
		\begin{aligned}
			\widetilde{T}_{(L)\,ab} &= \frac{2}{\sqrt{-h}}\frac{\delta}{\delta h^{ab}}(i\ln Z_{\text{dS}}) \\
			&= \frac{2}{\sqrt{-h}}\frac{\delta}{\delta h^{ab}}(S_{\text{dS}}^{\text{on-shell}}) \\
			&= -\frac{1}{8\pi G}\left(K_{ab} - K h_{ab} + \frac{h_{ab}}{\lds}\right).
		\end{aligned}
	\end{equation}
	For the timelike boundary, performing a $2+1$ decomposition along the radial direction and substituting $n_a n^a = 1$ into the Hamiltonian constraint~\eqref{WdWequation} yields
	\begin{equation}
		\widetilde{\mathcal{R}}^{(2)} - K^2 + K_{ab}K^{ab} - 2\Lambda_2 = 0.
	\end{equation}
	Following the same procedure as in the previous subsection, one finally obtains the trace flow equation when the dual field theory resides on the timelike boundary $\partial\mathcal{M}$,
	\begin{equation}
		\widetilde{T}_{(L)}=\frac{\lds}{16\pi G}\widetilde{\mathcal{R}}^{(2)} + 4\pi G\lds\left(\widetilde{T}_{(L)}^{ab}\widetilde{T}_{(L)\,ab}- \widetilde{T}_{(L)}^2\right) - \frac{1}{4\pi G\lds}.
	\end{equation}
	Neglecting the trace anomaly and employing the duality relation~\eqref{dsconstrant}, the contribution of the deformation operator to the trace of the boundary stress tensor can be expressed as
	\begin{equation}
		T_{(L)} = \frac{\pi\lambda_{\text{dS}}}{2}\left(T_{(L)}^{ab}T_{(L)\,ab} - T_{(L)}^2\right) - \frac{2}{\pi\lambda_\ds}.
	\end{equation}
	Considering the variation of the only scale in the boundary theory from $\lambda_{\ds}$ to $\lambda_{\ds}+\delta \lambda_{\ds}$, the infinitesimal variation of the physical boundary metric is then given by\footnote{In contrast to the case with a spacelike boundary, the deformation parameter in the $T\bar{T}+\Lambda_2$ deformation is real, $\lambda_{\text{dS}}=\frac{8G\lds}{r_c^2}$.}
	\begin{equation}
		\delta\gamma_{ab} = -\frac{\delta\lambda_{\ds}}{\lambda_{\ds}}\gamma_{ab}, \quad 
		\delta\gamma^{ab} = \frac{\delta\lambda_{\ds}}{\lambda_{\ds}}\gamma^{ab}.
	\end{equation}
	Using the definition of the stress tensor in Eq.~\eqref{eq:BY define timelike}, one finds the variation of the dual field theory partition function on the timelike boundary:
	\begin{equation}
			\delta(i\ln Z_{\text{QFT}})=\frac{1}{2}\int d^2x \sqrt{-\gamma}\,T_{(L)\,ab}\delta\gamma^{ab}
			=\frac{1}{2\lambda_\ds}\int d^2x \sqrt{-\gamma}\,T_{(L)}\,\delta\lambda_\ds.
	\end{equation}
	Consequently, the flow equation for the partition function on the field theory side can be obtained as
	\begin{equation}
		\partial_{\lambda_\ds} (i\ln Z_{\text{QFT}})= \frac{1}{2\lambda_\ds}\int d^2x \sqrt{-\gamma}\,T_{(L)}.
	\end{equation}
	In terms of the action, this flow equation becomes
	\begin{equation}
		\label{the second step}
			\partial_{\lambda_\ds}S_{\text{QFT}} = \frac{\pi}{4} \int d^2x\,\sqrt{-\gamma}\,\left(T_{(L)}^{ab}T_{(L)\,ab} - T_{(L)}^2\right)
			- \frac{1}{\pi\lambda_\ds^2}\int d^2x\,\sqrt{-\gamma}.
	\end{equation}

	\section{Energy spectrum }
	\label{Energy spectrum}

	In this section, we compute separately the dependence of the energy spectrum on the deformation parameter for the $T\bar{T}$ and $T\bar{T}+\Lambda_2$ deformations. By imposing the continuity condition on the energy, the spectra in both cases are matched at the critical deformation point where the spectrum becomes complex, thereby preventing the energy spectrum from becoming complex and allowing the deformation parameter to extend continuously to the infinite.

		\subsection{The first step: $T\bar{T}$ deformation}

\begin{figure}[H]
		\centering
		\includegraphics[scale=0.6]{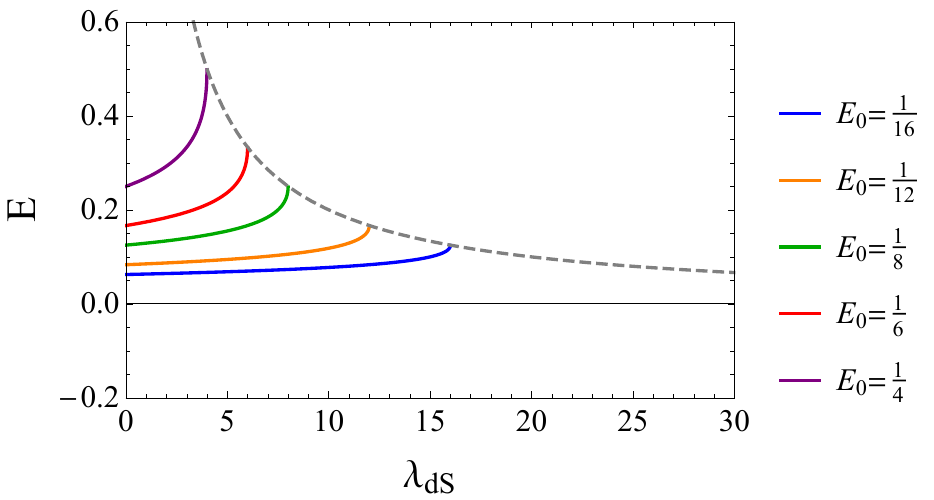}
		\caption{
			The energy flow of the $T\bar{T}$ deformation with $L=2\pi$ and $J=0$. One finds that when the deformation parameter $\lambda_{\text{dS}}$ exceeds $\frac{1}{E_0}$, the energy becomes complex.}
		\label{first step}
	\end{figure}

	We solve the flow equation for the energy spectrum by placing the dual field theory on a cylindrical manifold. For the $T\bar{T}$ deformation with the spacelike boundary, the physical boundary metric is chosen to be Euclidean,
	\begin{equation}
		ds^2 = dt^2 + d\phi^2.
	\end{equation}
	In the Euclidean framework, the relation between the interacting Hamiltonian $\mathcal{H}_{\text{int}}$ and the interacting Lagrangian density $\mathcal{L}_{\text{int}}$ is given by $\mathcal{H}_{\text{int}}=\int d\phi\, \mathcal{L}_{\text{int}}$. Since we adopt the convention $Z_{\text{dS}}=\int \mathcal{D}\psi\, e^{-iS_{\text{dS}}}$ for the partition function, the corresponding relation in our notation becomes
	\begin{equation}
		\mathcal{H}_{\text{int}}=-\int d\phi\, \mathcal{L}_{\text{int}}.
	\end{equation}
	Expanding the flow equation~\eqref{the first step} in the Euclidean signature gives
	\begin{equation}
		\partial_{\lambda_\ds}S_{\text{QFT}} =-\frac{\pi}{2}\int dtd\phi(T_{(E)\,\phi\phi}T_{(E)\,tt}-T_{(E)\,t\phi}^2).
	\end{equation}
	Hence, the corresponding flow equation for the Hamiltonian reads
	\begin{equation}
		\partial_{\lambda_\ds}\mathcal{H} =\frac{\pi }{2}\int d\phi(T_{(E)\,\phi\phi}T_{(E)\,tt}-T_{(E)\,t\phi}^2).
	\end{equation}
	For an energy eigenstate $|E,J\rangle$, the derivative of the energy with respect to the deformation parameter $\lambda_{\ds}$ is obtained from the Feynman--Hellmann theorem as
	\begin{equation}
		\begin{aligned}
			\partial_{\lambda_{\ds}} E
			&=\langle E,J| \partial_{\lambda_\ds}\mathcal{H} |E,J\rangle \\
			&=\frac{\pi }{2}\int d\phi\,\langle E,J|T_{(E)\,\phi\phi}T_{(E)\,tt}-T_{(E)\,t\phi}^2|E,J\rangle \\
			&=\frac{\pi L}{2}\,(\langle T_{(E)\,\phi\phi}\rangle\langle T_{(E)\,tt}\rangle-\langle T_{(E)\,t\phi}^2\rangle),
		\end{aligned}
	\end{equation}
	where $L=\int d\phi=2\pi$ denotes the circumference of the cylinder. According to the definition of the Brown--York tensor in Eq.~\eqref{eq:BY define}, the dual stress tensor carries an additional imaginary factor $i$, and its components are therefore related to the energy, momentum, and angular momentum as~\cite{AliAhmad:2025kki}
	\begin{equation}
		\label{enerygexplictform}
		\langle T_{(E)\,tt}\rangle=i\frac{ E}{L},\quad \langle T_{(E)\,\phi\phi}\rangle=i\frac{\partial E}{\partial L},\quad \langle T_{(E)\,t\phi}\rangle=i\frac{2\pi J}{L^2}.
	\end{equation}
	The energy flow equation is then obtained as
	\begin{equation}
		\frac{2}{\pi}\partial_{\lambda_{\ds}} E=-E\frac{\partial E}{\partial L}+\frac{4\pi^2 J^2}{L^3}.
	\end{equation}
	To solve this equation, we introduce the dimensionless energy $\mathcal{E}(\lambda_{\ds}/L^2)=L E$, which enables an explicit solution by variable substitution,
	\begin{equation}
		\label{deformed dimensionless energy first step}
		\mathcal{E}=\frac{L^2}{\pi\lambda_{\ds}}\left(1-\sqrt{1-\frac{2\pi\lambda_{\ds} \mathcal{E}_0}{L^2}-\frac{4J^2\pi^4\lambda_{\ds}^2}{L^4}}\right),
	\end{equation}
	where $\mathcal{E}_0=2\pi(\Delta-\frac{c}{12})$ denotes the dimensionless energy of the seed CFT. The corresponding dressed energy is
	\begin{equation}
		\label{deformed energy first step}
		E=\frac{L}{\pi\lambda_{\ds}}\left(1-\sqrt{1-\frac{2\pi\lambda_{\ds} E_0}{L}-\frac{4J^2\pi^4\lambda_{\ds}^2}{L^4}}\right),
	\end{equation}
	with $E_0=\frac{\mathcal{E}_0}{L}$ representing the undeformed energy. As illustrated in Fig.~\ref{first step}, for different initial energy levels, the energy inevitably develops a complex value as the deformation parameter increases. The critical deformation parameter $\lambda_0$ at which the energy becomes complex is\footnote{The other solution for the critical deformation parameter is discarded, since it takes a negative value.}
	\begin{equation}
		\lambda_0=\frac{E_0L^3(\sqrt{1+\frac{4J^2\pi^2}{E_0^2L^2}}-1)}{4J^2\pi^3}=\frac{2(\sqrt{E_0^2+J^2}-E_0)}{J^2},
	\end{equation}
	where we have taken $L=2\pi$ in the last step. For the case of $J=0$, the critical deformation parameter is given by $\lambda_0=\frac{1}{E_0}$.

	\subsection{The second step: $T\bar{T}+\Lambda_2$ deformation\label{The second step}}
	
\begin{figure}[H]
		\centering
		\includegraphics[scale=0.6]{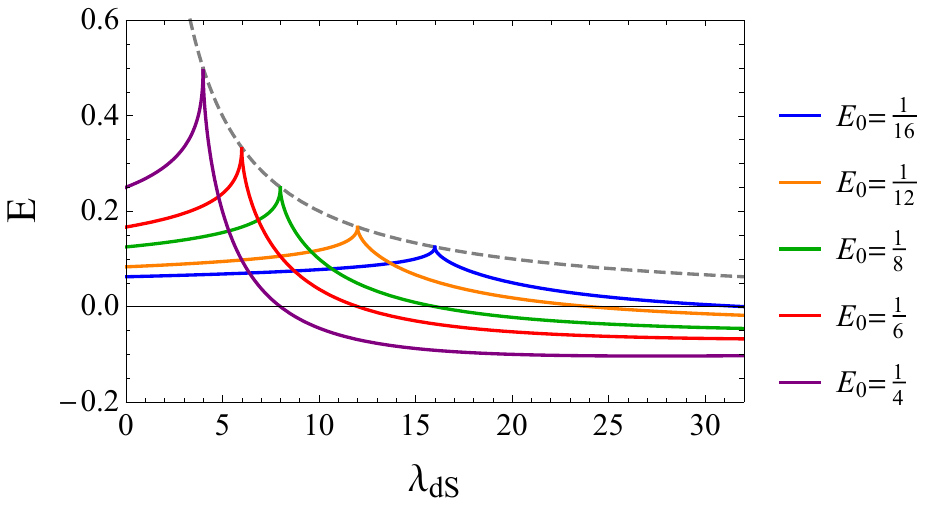}
		\caption{Energy flow of the $T\bar{T}+\Lambda_2$ deformation with $L=2\pi$ and $J=0$.}
		\label{second step}
	\end{figure}

	To prevent the emergence of complex energy eigenvalues, when the deformation parameter attains its critical value $\lambda_0$, the subsequent deformation is governed by the $T\bar{T}+\Lambda_2$ deformation. In this regime, the dual field theory is defined on a manifold with Lorentzian signature. The explicit form of the corresponding flow equation~\eqref{the second step} takes the form
	\begin{equation}
		\partial_{\lambda_\ds}S_{\text{QFT}} = \frac{\pi}{2}\int dtd\phi\left(T_{(L)\,\phi\phi}T_{(L)\,tt}-T_{(L)\,t\phi}^2-\frac{2}{\pi^2\lambda_{\ds}^2}\right).
	\end{equation}
	Under Lorentzian signature, and according to our convention for the partition function, the relation between the interacting Hamiltonian and the Lagrangian density is given by
	\begin{equation}
		\mathcal{H}_{\text{int}}=\int d\phi\, \mathcal{L}_{\text{int}}.
	\end{equation}
	Therefore, the corresponding energy flow equation becomes
	\begin{equation}
		\partial_{\lambda_{\ds}} E=\frac{\pi L}{2}\left(\langle T_{(L)\,\phi\phi}\rangle\langle T_{(L)\,tt}\rangle-\langle T_{(L)\,t\phi}^2\rangle\right)-\frac{L}{\pi\lambda_{\ds}^2},
	\end{equation}
	where the expectation values of the stress tensor are related to the energy $E$ and angular momentum $J$ through
	\begin{equation}
		\langle T_{(L)\,tt}\rangle=\frac{E}{L},\quad \langle T_{(L)\,\phi\phi}\rangle=-\frac{\partial E}{\partial L},\quad \langle T_{(L)\,t\phi}\rangle=-\frac{2\pi J}{L^2}.
	\end{equation}
	Consequently, the energy flow equation can be expressed as
	\begin{equation}
		\label{energy flow equation second step}
		\frac{2}{\pi}\partial_{\lambda_{\ds}} E=-E\frac{\partial E}{\partial L}-\frac{4\pi^2 J^2}{L^3}-\frac{2L}{\pi^2\lambda_{\ds}^2}.
	\end{equation}
	By introducing the dimensionless energy $\mathcal{E}=EL$ and imposing the continuity condition at the critical deformation parameter,
	\begin{equation}
		E_{\lambda_0}^{(T\bar{T})}=E_{\lambda_0}^{(T\bar{T}+\Lambda_2)},
	\end{equation}
	the solution to the energy flow equation~\eqref{energy flow equation second step} is obtained as
	\begin{equation}
		\label{deformed energy second step}
		E=\frac{L}{\pi\lambda_{\ds}}\left(1-\sqrt{-1+\frac{2\pi\lambda_{\ds} E_0}{L}+\frac{4J^2\pi^4\lambda_{\ds}^2}{L^4}}\right).
	\end{equation}
	As shown in Fig.~\ref{second step}, matching the $T\bar{T}$ and $T\bar{T}+\Lambda_2$ flows at the critical deformation parameter $\lambda_0$  avoids the complexification of the energy spectrum, allowing the deformation parameter to continuously approach infinity. We emphasize that, when the deformation parameter is increased to the critical value $\lambda_0$, we switch from the $T\bar T$ deformation to the $T\bar T+\Lambda_2$ deformation, while the signature of the manifold on which the field theory is defined changes from Euclidean to Lorentzian. This is because the $T\bar T$-deformed theory is defined on a spacelike surface, whereas the $T\bar T+\Lambda_2$-deformed theory is defined on a timelike surface. In the following subsection, the flows of the metric and the stress tensor will be discussed in more detail.

	\subsection{Variational methods for solving energy spectrum}
	Beyond the standard approach of determining the energy spectrum via the flow equation for the action, one may also obtain the spectrum by employing a variational principle~\cite{Guica:2019nzm}. Furthermore, the flow of metric can be made manifest in this approach.  We will take the $T\bar{T}$ deformation as an example to illustrate this method. Using the variational form \eqref{variation of Z} and the differential form \eqref{partition flow equation} of the partition function, one finds
	\begin{equation}
		-\frac{1}{2}\partial_\lambda\left(\sqrt{\gamma}T_{(E)\,ab}\delta\gamma^{ab}\right)
		=\frac{\pi}{4}\delta\left[\sqrt{\gamma}\left(T_{(E)}^{ab}T_{(E)\,ab} - T_{(E)}^2\right)\right].
	\end{equation}
	Taking into account that the deformation parameters are related by $\lambda=-i\lambda_{\ds}$, and redefining the stress tensor as $T^{\prime}_{(E)\,ab}=\frac{\pi}{4}T_{(E)\,ab}$ for convenience, the above expression simplifies to
	\begin{equation}
		\partial_{\lambda_{\ds}}\left(\sqrt{\gamma}T^{\prime}_{(E)\,ab}\delta\gamma^{ab}\right)
		=2i\delta\left[\sqrt{\gamma}\left(T^{\prime\,ab}_{(E)}T^{\prime}_{(E)\,ab} - T^{\prime\,2}_{(E)}\right)\right].
	\end{equation}
	This expression can be decomposed as
	\begin{equation}
		\partial_{\lambda_{\ds}}\gamma_{ab}=4i\hat{T}_{(E)\,ab},\qquad
		\partial_{\lambda_{\ds}}\hat{T}_{(E)\,ab}=2i\hat{T}_{(E)\,ac}\hat{T}^{\quad c}_{(E)\,b},
	\end{equation}
	where $\hat{T}_{(E)\,ab}=\left(T^{\prime}_{(E)\,ab}-\gamma_{ab}T_{(E)}^{\prime}\right)$. To avoid the computational complications arising from explicit factors of the imaginary unit $i$, we further redefine the stress tensor by setting $\ddot{T}^{\prime}_{(E)\,ab}=iT^{\prime}_{(E)\,ab}$. This leads to decoupled equations that take the same form as those obtained in Ref.~\cite{AliAhmad:2025kki},
	\begin{equation}
		\partial_{\lambda_{\ds}}\gamma_{ab}=4\hat{T}^{\prime}_{(E)\,ab},\qquad
		\partial_{\lambda_{\ds}}\hat{T}^{\prime}_{(E)\,ab}=2\hat{T}^{\prime}_{(E)\,ac}\hat{T}^{\prime\quad c}_{(E)\,b},
	\end{equation}
	where $\hat{T}^{\prime}_{(E)\,ab}=\left(\ddot{T}^{\prime}_{(E)\,ab}-\gamma_{ab}\ddot{T}_{(E)}^{\prime}\right)$. Consequently, the metric flow is given by
	\begin{equation}
		\gamma_{ab}
		=\gamma_{ab}^{[0]}+4\lambda_{\ds}\hat{T}^{\prime\,[0]}_{(E)\,ab}
		+4\lambda_{\ds}^2\hat{T}^{\prime\,[0]}_{(E)\,ac}\hat{T}^{\prime\,[0]}_{(E)\,bd}\gamma^{cd\,[0]},
	\end{equation}
	while the stress tensor flow takes the form
	\begin{equation}
		\label{stress tensor flow}
		\hat{T}^{\prime}_{(E)\,ab}
		=\hat{T}^{\prime\,[0]}_{(E)\,ab}
		+2\lambda_{\ds}\hat{T}^{\prime\,[0]}_{(E)\,ac}\hat{T}^{\prime\,[0]}_{(E)\,bd}\gamma^{cd\,[0]}.
	\end{equation}
	Here $\gamma_{ab}^{[0]}$ and $\hat{T}^{\prime\,[0]}_{(E)\,ab}$ denote the initial metric and the initial stress tensor at $\lambda_{\ds}=0$, respectively. In our convention, the initial metric is chosen to be the flat Euclidean metric $\gamma_{ab}^{[0]}=\delta_{ab}$. The deformed metric and the Euclidean metric are then related by a diffeomorphism,
	\begin{equation}
		\label{diffeomorphism relation}
		\begin{aligned}
			\gamma_{ab}
			&=\delta_{ab}
			+4\lambda_{\ds}\hat{T}^{\prime\,[0]}_{(E)\,ab}
			+4\lambda_{\ds}^2\hat{T}^{\prime\,[0]}_{(E)\,ac}\hat{T}^{\prime\,[0]}_{(E)\,bd}\delta^{cd} \\
			&=\left(\delta_{a}^{\ c}+2\lambda_{\ds}\hat{T}^{\prime\,[0]}_{(E)\,ae}\delta^{ec}\right)
			\delta_{cd}
			\left(\delta_{b}^{\ d}+2\lambda_{\ds}\hat{T}^{\prime\,[0]}_{(E)\,fb}\delta^{df}\right).
		\end{aligned}
	\end{equation}
	However, it should be emphasized that the $T\bar T$ deformation does not alter the form of the background manifold metric itself; its effect on the metric arises only through a Weyl factor that depends on the deformation parameter. In other words, the redefined stress tensor introduced here is defined solely within the coordinate system associated with the metric $\gamma_{ab}$, whereas the physical stress tensor is defined in the coordinates $y_{\lambda}$ in which the metric $\delta_{ab}$ remains Euclidean \cite{AliAhmad:2025kki}. More precisely,
	\begin{equation}
		\label{firstgauge}
		\gamma_{ab}dx^adx^b=\delta_{ab}dy^a_{\lambda}dy^b_{\lambda},\quad \frac{dy^c_{\lambda}}{dx^a}=\delta_{a}^{c}+2\lambda_{\ds}\hat{T}^{\prime\,[0]}_{(E)\,ae}\delta^{ec}.
	\end{equation}
	The relation between the physical stress tensor $\mathcal{T}_{ab}$ and the deformed stress tensor $\ddot{T}^{\prime}_{(E)\,ab}$ is then given by
	\begin{equation}
		\ddot{T}^{\prime}_{(E)\,ab}=\left(\delta_{a}^{c}+2\lambda_{\ds}\hat{T}^{\prime\,[0]}_{(E)\,ae}\delta^{ec}\right)\mathcal{T}_{cd}\left(\delta_{b}^{d}+2\lambda_{\ds}\hat{T}^{\prime\,[0]}_{(E)\,fb}\delta^{df}\right).
	\end{equation}
	Substituting Eqs.~\eqref{stress tensor flow} and \eqref{diffeomorphism relation} into the above expression and simplifying, one obtains
	\begin{equation}
		\ddot{T}^{\prime\,[0]}_{(E)\,ad}=\left(\delta_{a}^{c}+2\lambda_{\ds}\hat{T}^{\prime\,[0]}_{(E)\,ae}\delta^{ec}\right)\mathcal{T}_{cd}.
	\end{equation}
	Taking into account that the undeformed physical stress tensor $\mathcal{T}_{cd}^{[0]}$ is defined in the coordinates associated with $y_0^c$, with
	\begin{equation}
		\label{secondgauge}
		\delta_{ab}dx^adx^b=\Omega^2\delta_{cd}dy_0^cdy_0^d,
	\end{equation}
	where $\Omega$ denotes the Weyl factor depending on the deformation parameter, one finally obtains the dependence of the physical stress tensor on the deformation parameter as
	\begin{equation}
		\mathcal{T}_{cd}^{[0]}=\frac{dx^a}{dy_0^c}\left(\delta_{a}^{f}+2\lambda_{\ds}\hat{T}^{\prime\,[0]}_{(E)\,ae}\delta^{ef}\right)\mathcal{T}_{fb}\frac{dx^b}{dy_0^d}.
	\end{equation}
	In the above derivation, Eqs.~\eqref{firstgauge} and \eqref{secondgauge} introduce two $\mathrm{SO}(2)$ gauge redundancies. These redundancies must be fixed by imposing the spatial periodicity of the physical coordinates $y^a$ together with the specified undeformed physical stress tensor. One then obtains the individual components of the physical stress tensor as
	\begin{equation}
		\begin{aligned}
			\mathcal{T}_{tt}&=-\frac{1}{4\lambda_{\ds}}\left(1-\sqrt{1+8\lambda_{\ds}  \mathcal{T}_{tt}^{[0]}-16\lambda_{\ds}^2(\mathcal{T}_{tx}^{[0]})^2}\right)\,,\\
			\mathcal{T}_{tx}&=\mathcal{T}_{tx}^{[0]}\,,\\
			\mathcal{T}_{xx}&=-\frac{1-16\lambda_{\ds}^2(\mathcal{T}_{tx}^{[0]})^2-\sqrt{1+8\lambda_{\ds}  \mathcal{T}_{tt}^{[0]}-16\lambda_{\ds}^2(\mathcal{T}_{tx}^{[0]})^2}}{4\lambda_{\ds}\sqrt{1+8\lambda_{\ds}  \mathcal{T}_{tt}^{[0]}-16\lambda_{\ds}^2(\mathcal{T}_{tx}^{[0]})^2}}\,.
		\end{aligned}
	\end{equation}
	We now return to our original convention
	\begin{equation}
		\mathcal{T}_{ab}=i\frac{\pi}{4}T_{(E)\,ab}\,,\quad \mathcal{T}_{ab}^{[0]}=i\frac{\pi}{4}T_{(E)\,ab}^{[0]}\,.
	\end{equation}
	Moreover, taking into account the explicit form of the stress tensor components in the $T\bar T$ deformation given in Eq.~\eqref{enerygexplictform}, one finally obtains
	\begin{equation}
		\small
		\mathcal{T}_{tt}=-\frac{\pi}{4}\frac{E}{L}=-\frac{1}{4\lambda_{\ds}}\left(1-\sqrt{1-\frac{2\pi\lambda_{\ds} E_0}{L}-\frac{4J^2\pi^4\lambda_{\ds}^2}{L^4}}\right).
	\end{equation}
	This result is in complete agreement with Eq.~\eqref{deformed energy first step}. In addition, by using the metric flow, one can also derive the dependence of the determinant of the metric on the deformed manifold with respect to the deformation parameter as
	\begin{equation}
		\begin{aligned}
			\det[\gamma]&=\left(\det[I-2\lambda_{\ds}\delta^{-1}\hat{\mathcal{T}}_{(E)}]\right)^{-2}\\
			&=\left(\frac{2\sqrt{1+8\lambda_{\ds}  \mathcal{T}_{tt}^{[0]}-16\lambda_{\ds}^2(\mathcal{T}_{tx}^{[0]})^2}}{1+4\lambda_{\ds}\mathcal{T}_{tx}^{[0]}+\sqrt{1+8\lambda_{\ds}  \mathcal{T}_{tt}^{[0]}-16\lambda_{\ds}^2(\mathcal{T}_{tx}^{[0]})^2}}\right)^2\\
			&=\left(\frac{2\sqrt{1-\frac{2\pi\lambda_{\ds} E_0}{L}-\frac{4J^2\pi^4\lambda_{\ds}^2}{L^4}}}{1-\frac{\pi^2J\lambda_{\ds}}{2L^2}+\sqrt{1-\frac{2\pi\lambda_{\ds} E_0}{L}-\frac{4J^2\pi^4\lambda_{\ds}^2}{L^4}}}\right)^2\,.
		\end{aligned}
	\end{equation}
	At the critical value of the deformation parameter, the determinant of the metric on the deformed manifold vanishes, indicating that the manifold becomes a null hypersurface. In Sec.~\ref{The second step}, the transition of the deformation from the $T\bar T$ deformation to the $T\bar T+\Lambda_2$ deformation, together with the change in the signature of the manifold metric and the vanishing of the boundary metric determinant at the critical deformation parameter, is naturally reminiscent of the transition of constant-$r$ hypersurfaces in dS static coordinates from spacelike to timelike. In dS static coordinates, the radial coordinate $r$ is timelike outside the cosmological horizon and spacelike inside it. Consequently, the induced metric on the $r=\text{const.}$ hypersurface has Euclidean signature outside the horizon and Lorentzian signature inside. In particular, the cosmological horizon, being a null hypersurface, is characterized by a vanishing metric determinant. Motivated by this observation, we propose that, in dS holography, the composite flow formed by the $T\bar T$ flow and the $T\bar T+\Lambda_2$ flow is holographically dual to the motion of the boundary in dS static coordinates from infinity into the bulk, crossing the cosmological horizon and approaching the worldline of a static observer. This holographic proposal has the potential to unify two a priori distinct classes of dS holographic models. In the following, we test this proposal from the perspectives of quasi-local energy and holographic entanglement entropy.

	\section{Bulk analysis}
	\label{Bulk analysic}

	To explore the holographic dual of the above composite flow, we propose that the bulk geometry consists of the dS static patch and its extension, based on the correspondence between the signature transition of the two-dimensional manifold on which the dual field theory resides and the transition of the $r={\rm const.}$ hypersurfaces in the dS static coordinates from timelike to spacelike across the cosmological horizon. The first step of the composite flow, namely the $T\bar{T}$ deformation, corresponds to the inward shift of the future spacelike infinity in the extended region of dS spacetime. The extended dS static spacetime is given by
	\begin{equation}
		\label{extended coordinates}
		ds^2=-\frac{1}{F(r)}dr^2+F(r)dt^2+r^2d\phi^2, \quad F(r)=\frac{r^2}{\lds^2}-1>0.
	\end{equation}
	\begin{figure}[H]
		\centering
		\includegraphics[scale=0.8]{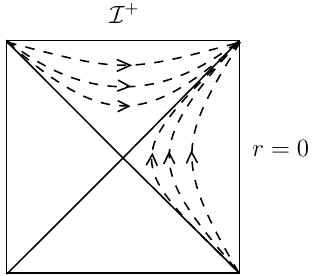}
		\caption{Killing vector field. The Killing vector $\xi^{\mu}$ is timelike in the static region and spacelike in the extended region.}
		\label{killing flow}
	\end{figure}
	The spacetime boundary is located at the spacelike infinity $r=\infty$. Although dS spacetime possesses neither a timelike boundary nor a globally defined timelike Killing vector field, one can still define a notion of energy by employing the boundary Brown--York tensor~\cite{Brown:1992br,Balasubramanian:1999re}, in analogy with that in AdS spacetime~\cite{Balasubramanian:2001nb}. For the dS static coordinates~\eqref{ds static coordinate} and the extended coordinates~\eqref{extended coordinates}, one can identify a Killing vector field $\xi^{\mu}$ that is timelike in the static region and spacelike in the extended spacetime, as shown in Fig.~\ref{killing flow}.

	For the spacelike boundary at $r=r_c$, the quasi-local energy is defined as~\cite{Balasubramanian:2001nb}
	\begin{equation}
		E=\int_0^{2\pi}d\phi\,\sqrt{g_{\phi\phi}}\,u^{\mu}\xi^{\nu}\tilde{T}_{\mu\nu},
	\end{equation}
	where $u^{\mu}$ denotes the unit normal vector to the curve of fixed $t$ on the boundary at $r=r_c$. Specifically,
	\begin{equation}
		u^t=\frac{1}{\sqrt{F(r)}}\,,\qquad u^{\phi}=0.
	\end{equation}
	With the normalized Killing vector field $\xi^{\mu}=\frac{1}{\sqrt{F(r)}}(\partial_t)^{\mu}$, the quasi-energy density is given by\footnote{The factor of $i$ in $\tilde{T}_{tt}$ has been canceled by that in the energy $iE$.}
	\begin{equation}
		u^{t}\xi^{t}\tilde{T}_{tt}=\frac{u^{t}\xi^{t}}{8\pi G}\left(K_{tt}-Kh_{tt}+\frac{h_{tt}}{\lds}\right)
		=\frac{r_c-\sqrt{r_c^2-\lds^2}}{8\pi G\lds r_c}.
	\end{equation}
	Integrating over the spatial section, the total energy of the spacetime becomes
	\begin{equation}
		E=\frac{r_c}{4G\lds}\left(1-\sqrt{1-\frac{\lds^2}{r_c^2}}\right)
		=\frac{r_c}{4G\lds}\left(1-\sqrt{1-\frac{8G\lds E_0}{r_c}}\right),
	\end{equation}
	where $E_0=\frac{\lds}{8Gr_c}$ denotes the undeformed energy of dS spacetime. The spatial proper length of the spacelike boundary is $L=\int_0^{2\pi}d\phi\,\sqrt{g_{\phi\phi}}=2\pi r_c$. Multiplying it by the total energy $E$ yields a dimensionless energy
	\begin{equation}
		\mathcal{E}=\frac{\pi r_c^2}{2G\lds}\left(1-\sqrt{1-\frac{8G\lds \mathcal{E}_0}{2\pi r_c^2}}\right),
	\end{equation}
	where $\mathcal{E}_0=L E_0=\frac{\pi\lds}{4G}$. This result precisely reproduces Eq.~\eqref{deformed dimensionless energy first step} on the field theory side with vanishing angular momentum $J=0$\footnote{Since the radial coordinate $r$ determines the proper circumference of the boundary, the dependence of the deformation parameter on $r$ disappears, i.e., $L=2\pi r_c$ and $\lambda_{\ds}=8G\lds$.}.

\begin{figure}[H]
		\centering
		\includegraphics[scale=0.6]{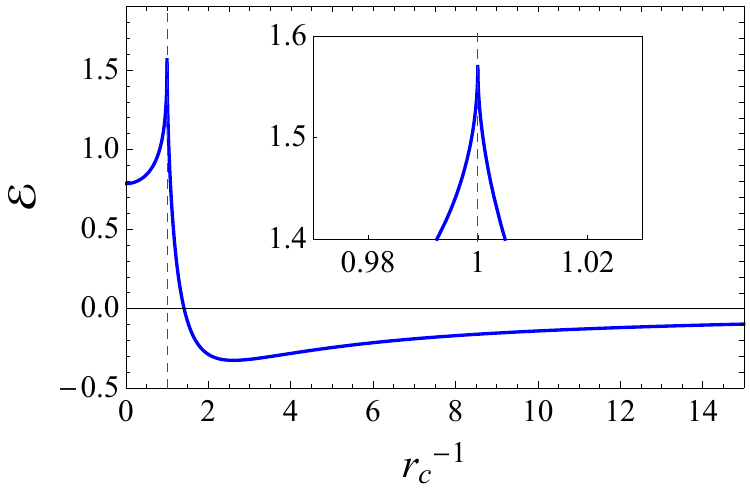}
		\caption{Dimensionless energy in dS spacetime. We take $G=\lds=1$. The red dashed line denotes the cosmological horizon $r=\lds$.}
		\label{bulkthrough}
	\end{figure}

	When the radial cutoff $r_c$ becomes smaller than $\lds$, the spacelike boundary crosses the cosmological horizon and turns into a timelike boundary. The conserved charge is then computed using the timelike Killing vector field in the static region. It should be noted that due to the change in the metric signature, the definition of the Brown--York tensor acquires an additional minus sign,
	\begin{equation}
		\tilde{T}_{tt}=-\frac{1}{8\pi G}\left(K_{tt}-Kh_{tt}+\frac{h_{tt}}{\lds}\right)
		=\frac{(\lds^2-r_c^2)(r_c-\sqrt{\lds^2-r_c^2})}{8\pi G\lds^3r_c}.
	\end{equation}
	Consequently, the total energy in the dS static spacetime with finite cutoff $r=r_c$ is given by
	\begin{equation}
		E=\frac{r_c}{4G\lds}\left(1-\sqrt{-1+\frac{\lds^2}{r_c^2}}\right)
		=\frac{r_c}{4G\lds}\left(1-\sqrt{-1+\frac{8G\lds E_0}{r_c}}\right).
	\end{equation}
	This result precisely matches the expression~\eqref{deformed energy second step} derived from the $T\bar{T}+\Lambda_2$ deformation with $J=0$. Finally, we obtain the complete energy flow in dS spacetime, as illustrated in Fig.~\ref{bulkthrough}.
	
	Furthermore, substituting the dimensionless seed energy $\mathcal{E}_0 = \frac{\pi\lds}{4G}$ into the Cardy formula~\cite{Cardy:1986ie,Mukhametzhanov:2019pzy}, we obtain
	\begin{equation}
		S_{\text{Cardy}} = 2\pi \sqrt{\frac{c_{\text{dS}}}{3}\left(\Delta-\frac{c_{\text{dS}}}{12}\right)}
		= 2\pi \sqrt{\frac{c_{\text{dS}}}{3}\frac{\mathcal{E}_0}{2\pi}}
		= \frac{\pi \lds}{2G}
		= S_{\text{dS}},
	\end{equation}
	where $S_{\text{dS}}=\frac{\pi \lds}{2G}$ is the dS entropy, defined as the area of the cosmological horizon divided by $4G$. This demonstrates that our holographic construction provides a microscopic accounting of the dS entropy. In contrast to Ref.~\cite{Coleman:2021nor}, which accounts for the dS entropy by considering the thermal state dual to a BTZ black hole at the special energy level $\Delta = \frac{c}{6}$, we instead work with the thermal state associated with the dS static spacetime to perform the microscopic counting directly. In addition, based on the dS$_3$/matrix integral duality, Ref.~\cite{Collier:2025lux} provided a microscopic description of the dS entropy by integrating the eigenvalue density. Meanwhile, Ref.~\cite{Wang:2025jfd} constructed a family of semi-classical dS microstates by considering the backreacted geometry of dS spacetime with a thin-shell brane, and obtained the same dS entropy through gravitational state counting.

	\section{Holographic Entanglement Entropy}
	\label{Holographic Entanglement Entropy}

	A central consistency check of holographic duality is that the holographic entanglement entropy matches the area of the Ryu--Takayanagi (RT) surface in the bulk divided by $4G$~\cite{Ryu:2006bv,Hubeny:2007xt}. In this section, we first review the general expression for the entanglement entropy of the dual field theory in dS spacetime using the spherical partition function method~\cite{Donnelly:2018bef,Lewkowycz:2019xse,Grieninger:2019zts,Chen:2023eic,Li:2020zjb}, building on our previous analysis of the $T\bar{T}$ deformation of entanglement entropy~\cite{Chang:2024voo}\footnote{For further developments on entanglement entropy under the $T\bar{T}$ deformation, see Refs.~\cite{Chen:2018eqk,He:2019vzf,Wang:2018jva,He:2020qcs, He:2022jyt,He:2023xnb, Apolo:2023ckr,PhysRevD.108.126013,Jeong:2022jmp,He:2023obo,He:2023wko,Deng:2023pjs,Wang:2024jem,Basu:2024enr,Basu:2024xjq,Basu:2025fsf,Lai:2025thy}.
	}. We then compute the entanglement entropy and the corresponding RT surface areas for specific subregions, both within the static patch and in the extended regions. This allows us to test the consistency of the unified holographic framework proposed in this work.

	\subsection{Entanglement entropy in dS spacetime}
	For an arbitrary two-dimensional manifold, the Casini--Huerta--Myers (CHM) mapping~\cite{Casini:2011kv} allows one to map the entangling region to a meridian at $\Phi = \mathrm{const.}$ on a replicated sphere,
	\begin{align}
		ds^2 = r^2 \Omega^2(\Theta, \Phi)\left(d\Theta^2 + n^2 \sin^2\Theta\, d\Phi^2\right).
		\label{eq:metric-conformalsphere}
	\end{align}
	In this setup, the entanglement entropy may be extracted from the endpoint contributions~\cite{Lewkowycz:2019xse,Chang:2024voo},
	\begin{align}
		r S'(r)
		= \lim_{n \rightarrow 1} 2\pi r^2 
		\left( 
		\Omega^2(0, \Phi)\!\!\int_{\Theta \in B(0, \epsilon)}
		+
		\Omega^2(\pi, \Phi)\!\!\int_{\Theta \in B(\pi, \epsilon)}
		\right)
		\sin\Theta\, d\Theta\, \partial_n \mathrm{tr}\,T,
		\label{eq:dC-conformalsphere}
	\end{align}
	where $B(p,\epsilon)$ denotes a small neighborhood of radius $\epsilon$ centered at $p$. Solving $\mathrm{tr}\,T$ from the trace flow equation~\eqref{eq:traceflowequ-CFT}, together with stress tensor conservation $\nabla_a T^{a}{}_b=0$, yields~\cite{Donnelly:2018bef,Chang:2024voo}
	\begin{align}
		T^{\Theta}_\Theta
		&=\frac{i}{\pi\lambda_\ds}
		\left(1-\sqrt{1-\frac{\lambda_\ds c_\ds}{12r_\eff^2} - \frac{\lambda_\ds c_\ds}{12r_\eff^2}\left(\frac{1}{n^2}-1\right)\frac{1}{\sin^2\Theta}}\right),\nn\\
		T^{\Phi}_\Phi
		&=\frac{i}{\pi\lambda_\ds}
		\left(1-\frac{1-\frac{\lambda_\ds c_\ds}{12r_\eff^2}}{\sqrt{1-\frac{\lambda_\ds c_\ds}{12r_\eff^2}-\frac{\lambda_\ds c_\ds}{12r_\eff^2}\left(\frac{1}{n^2}-1\right)\frac{1}{\sin^2\Theta}}}\right)\,,
		\label{eq:T-nsphere-dS}
	\end{align}
	where $r_{\eff}=r\Omega(\Theta,\Phi)$. Substituting these expressions into~\eqref{eq:dC-conformalsphere} and integrating over $r$, one obtains the general expression of the entanglement entropy:
	\begin{equation}
		\label{general entanglement entropy}
		S=-\frac{ic_\ds}{6}\tanh^{-1}\left(\frac{r_{\eff, \Theta = 0}}{\sqrt{r_{\eff, \Theta = 0}^2-\lds^2}}\right)-\frac{ic_\ds}{6}\tanh^{-1}\left(\frac{r_{\eff, \Theta = \pi}}{\sqrt{r_{\eff, \Theta = \pi}^2-\lds^2}}\right)\,.
	\end{equation}

	For the extended spacetime of the dS static coordinate, the induced metric on the boundary at $r=r_c$ is given by
	\begin{equation}
		ds^2=r_c^2\left(d\tilde{t}^2+d\phi^2\right),\quad \tilde{t}=\left(\frac{1}{\lds^2}-\frac{1}{r_c^2}\right)t.
	\end{equation}
	We choose the subregion $A \in [-\phi_a,\phi_a]$ at $t=0$ and compute its entanglement entropy. To proceed, we first map the cylindrical surface to the complex plane,
	\begin{align}
		\tilde{t} \rightarrow \frac{1}{2} \ln (T^2 + X^2)\,, \quad
		\phi \rightarrow \frac{1}{2i} \ln \left( \frac{T + iX}{T - iX} \right)\,.
		\label{eq:cyl-to-plane}
	\end{align}
	Applying the above coordinate transformation, we obtain the corresponding conformally flat metric,
	\begin{align}
		ds^2_{\mathrm{pl}} = r_c^2\frac{dT^2 + dX^2}{T^2 + X^2}\,.
	\end{align}
	We then use the CHM mapping to transform the conformally flat manifold into a conformal spherical manifold, under the coordinate transformation
	\begin{align}
		T\rightarrow\frac{\sin \phi_a\sin\Theta \sin \Phi}{1+\sin\Theta\cos\Phi}+\cos \phi_a\,,\quad
		X\rightarrow\frac{\sin \phi_a\cos\Theta}{1+\sin\Theta\cos\Phi}\,.
	\end{align}
	The resulting metric on the conformal sphere is
	\begin{align}
		ds^2_{\Omega\mathbb{S}^2}
		&=\frac{r_c^2\sin^2\phi_a}{(1+\cos(2\phi_a-\Phi)\sin\Theta)(1+\cos\Phi\sin\Theta)}\left(d\Theta^2+\sin^2\Theta d\Phi^2\right)\nn\\
		&=\Omega^2(\Theta,\Phi)r_\eff^2\left(d\Theta^2+\sin^2\Theta d\Phi^2\right)\,,
	\end{align}
	where $r_{\text{eff}}=r_c\sin\phi_a$. One observes that the conformal factor $\Omega(\Theta,\Phi)$ equals $1$ at the endpoints $\Theta=0$ and $\Theta=\pi$ of the subregion. Consequently, the entanglement entropy of the subregion $A$ is
	\begin{equation}
		\label{general entanglement entropy calculate}
		S=-\frac{ic_\ds}{3}\tanh^{-1}\left(\frac{r_{\eff}}{\sqrt{r_{\eff}^2-\lds^2}}\right)\,.
	\end{equation}
	For the case of $r_{\eff}>\lds$, the argument of the inverse hyperbolic tangent exceeds 1, so that $\tanh^{-1}$ acquires a complex value\footnote{For $x>1$, $\tanh^{-1}x=\coth^{-1}x+\frac{i\pi}{2}$.}. The entanglement entropy thus takes the form
	\begin{equation}\label{psuEE}
		S= -\frac{i c_\ds}{3} \coth^{-1} \left(\frac{r_{\eff}}{\sqrt{r_{\eff}^2-\lds^2}}\right) + \frac{\pi c_\ds}{6}.
	\end{equation}
	For the case of $r_{\eff}<\lds$, the argument under the square root becomes negative, and performing an appropriate analytic continuation yields
	\begin{equation}
		\label{finite entropy}
		S=\frac{c_\ds}{3}\tan^{-1}\left(\frac{r_{\eff}}{\sqrt{\lds^2-r_{\eff}^2}}\right).
	\end{equation}
	Now let us apply these general results to the dS static spacetime and its extended regions,
	\begin{equation}
		\label{staticspacetimemetric}
		ds^2=-\left(1-\frac{r^2}{\lds^2}\right)dt^2+\left(1-\frac{r^2}{\lds^2}\right)^{-1}dr^2+r^2d\phi^2,
	\end{equation}
	where $r\in(0,\lds)$ corresponds to the static patch, and $r\in(\lds,\infty)$ describes the extended spacetime region.

	\subsection{Extended spacetime: $r_{c}>\lds$}
	
	For the extended spacetime, we introduce the coordinate transformation $r=\lds\cosh\omega$ and $t=\lds \tau$, under which the line element becomes
	\begin{equation}
		\label{extened spacetime}
		ds^2=\lds^2\left(-d\omega^2+\sinh^2\omega \, d\tau^2+\cosh^2\omega \, d\phi^2 \right).
	\end{equation}
	In this description, the dual field theory resides on a two-dimensional Euclidean manifold located at a fixed value $\omega=\omega_c$. We consider the boundary subregion $A=[-\phi_a,\phi_a]$ at $\tau=0$. The corresponding effective radius is then
	\begin{equation}
		\label{extenedradial}
		r_{\text{eff}}=r_c\sin\phi_a=\lds\cosh\omega_c\sin\phi_a.
	\end{equation}
	Moreover, we require $\cosh\omega_c\sin\phi_a>1$ to ensure that the RT surface anchored to the boundary subregion $A$ lies entirely within the bulk spacetime, as illustrated in Fig.~\ref{geodesic}. The critical condition $\cosh\omega_c\sin\phi_a=1$ will be derived explicitly in the bulk calculation below. Finally, using the spherical partition function method, the entanglement entropy of the boundary subregion $A$ for $r_{c}>\lds$ is given by
	\begin{equation}
		\label{thefirstresult}
		S= -\frac{i c_\ds}{3} \coth^{-1} \left(\frac{\cosh\omega_c\sin\phi_a}{\sqrt{\cosh^2\omega_c\sin^2\phi_a-1}} \right) + \frac{\pi c_\ds}{6}.
	\end{equation}
	This expression takes precisely the form of the pseudo entropy in the dS/CFT correspondence, indicating that the density matrix of subsystem $A$ becomes non-Hermitian and thereby reflects the intrinsic non-unitarity of the dual field theory~\cite{Doi:2022iyj,Hikida:2022ltr}. This expression has been confirmed holographically through complex extremal surfaces in dS spacetime~\cite{Narayan:2015vda,Narayan:2022afv,Narayan:2023zen}.

	\subsubsection*{Bulk calculation:}
	
\begin{figure}[H]
		\centering
		\subfigure[$\cosh\omega_c\sin\phi_a<1$.]{
			\includegraphics[scale=1.0]{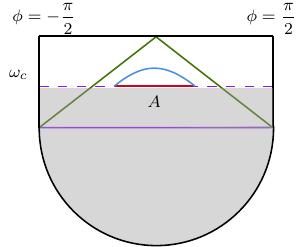}
			\label{geodesic1}}
		\hspace{1in}
		\subfigure[$\cosh\omega_c\sin\phi_a>1$.]{
			\includegraphics[scale=1.0]{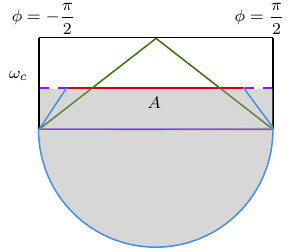}
			\label{geodesic2}
		}
		\caption{Schematic illustration of the RT surface corresponding to the boundary subregion $A$ in the extended spacetime. The purple solid line denotes the cosmological horizon $\omega=0$, while the purple dashed line corresponds to the spacelike boundary $\omega=\omega_c$. The dark green solid line represents a null geodesic, and the red segment indicates the boundary subregion $A$. The blue solid line represents the RT surface connecting the endpoints of subregion $A$. The shaded region corresponds to the bulk interior in the extended spacetime. \ref{geodesic1} When $\cosh\omega_c\sin\phi_a<1$, the RT surface associated with the subregion $A$ lies outside the bulk spacetime, and this configuration is excluded. \ref{geodesic2} When $\cosh\omega_c\sin\phi_a>1$, the RT surface associated with the boundary subregion $A$ is located inside the bulk spacetime, where the timelike geodesic segment outside the cosmological horizon contributes to the imaginary part of the entanglement entropy, while the Euclidean semicircular geodesic contributes to its real part.
		}
		\label{geodesic}
	\end{figure}

	For the extended spacetime~\eqref{extened spacetime}, we select the endpoints of the boundary subregion as $(\omega,\tau,\phi)=(\omega_c,0,-\phi_a)$ and $(\omega_c,0,\phi_a)$. By symmetry considerations, the geodesic lies within the $(\omega,\phi)$ plane, and its length can be computed by integrating the line element
	\begin{equation}
		D=\lds\int_{-\phi_a}^{\phi_a}d\phi\sqrt{\cosh^2\omega-\left(\frac{d\omega}{d\phi}\right)^2}.
	\end{equation}
	By extremizing the geodesic length, one obtains the corresponding differential equation
	\begin{equation}
		\frac{d\omega}{d\phi}=\cosh\omega\sqrt{1-\frac{\cosh^2\omega}{\cosh^2\omega_*}},
	\end{equation}
	where $\omega_*$ denotes the turning point of the geodesic as it extends into the bulk, corresponding to $\phi=0$. By performing the integration, one obtains
	\begin{equation}
		\phi_a=\frac{\pi}{2}-\tan^{-1}\left(\frac{\cosh\omega_*\sinh\omega_c}{\sqrt{\cosh^2\omega_*-\cosh^2\omega_c}}\right).
	\end{equation}
	When the turning point $\omega_*$ approaches spacelike infinity $\omega_*\to \infty$, $\phi_a$ takes its critical value
	\begin{equation}
		\phi_c=\frac{\pi}{2}-\tan^{-1}(\sinh\omega_c)=\sin^{-1}\left(\frac{1}{\cosh\omega_c}\right),
	\end{equation}
	and the geodesic connecting the endpoints of the boundary subregion $A$ becomes a null geodesic, as indicated by the dark green curve in Fig.~\ref{geodesic1}. 
	
	When $\phi_a<\phi_c$, the RT surface corresponding to the subregion $A$ is represented by the blue curve in Fig.~\ref{geodesic1}. In our holographic construction for the extended spacetime, the bulk region corresponds to $\omega<\omega_c$, and thus to ensure that the RT surface lies within the bulk, we require $\phi_a>\phi_c$, as illustrated in Fig.~\ref{geodesic2}. This condition guarantees that the effective spherical radius given in Eq.~(\ref{extenedradial}) is larger than $\lds$. In this case, the RT surface connecting the endpoints of subregion $A$ consists of timelike and spacelike geodesic segments, contributing respectively to the imaginary and real parts of the RT surface area. For the geodesic shown in Fig.~\ref{geodesic2}, it is convenient to compute its length using embedding coordinates. The three-dimensional dS spacetime can be considered as a hyperboloid in a four-dimensional Minkowski spacetime, with the constraint equation
	\begin{equation}
		-T^2+X^2+Y^2+W^2=\lds^2,
	\end{equation}
	where the uppercase letters $(X, Y, T, W)$ denote the embedding coordinates, and their relation to the extended spacetime coordinates $(\omega, \tau, \phi)$ is given by
	\begin{equation}
		\begin{aligned}
			&X=\lds\cosh \omega \cos \phi,\quad T=i\lds\sinh \omega \sinh \tau,\\
			&Y=\lds\cosh \omega \sin \phi,\quad W=i\lds \sinh \omega \cosh \tau.
		\end{aligned}
	\end{equation}
	In the four-dimensional Minkowski spacetime, the geodesic distance between points ``1'' and ``2'' is given by
	\begin{equation}
		\begin{aligned}
			D_{12}&=\lds \cos^{-1} \left(-T_{1} T_{2}+X_{1} X_{2}+Y_{1} Y_{2}+W_{1} W_{2}\right)\\
			&=\lds \cos^{-1} \left( 1-2\cosh^2\omega_c\sin^2\phi_a\right).	
		\end{aligned}
	\end{equation}
	Considering $\cosh\omega_c\sin\phi_a>1$, which leads to the argument of the inverse cosine function being less than $-1$, the geodesic distance can be obtained by employing trigonometric relations as\footnote{For the case of $x<-1$, $\cos^{-1}(x)=\pi-i\cosh^{-1}(-x)$.}
	\begin{equation}
		D_{12}= -i\lds\cosh^{-1} \left(2\cosh^2\omega_c\sin^2\phi_a-1\right)+\pi\lds.
	\end{equation}
	Dividing the above geodesic distance by $4G$ and using the dual relation $c_\ds=\frac{3\lds}{2G}$, one can verify that $\frac{D_{12}}{4G}$ precisely matches Eq.~(\ref{thefirstresult}):
	\begin{equation}
		\frac{D_{12}}{4G}=-\frac{i c_\ds}{3} \coth^{-1} \left(\frac{\cosh\omega_c\sin\phi_a}{\sqrt{\cosh^2\omega_c\sin^2\phi_a-1}}\right) + \frac{\pi c_\ds}{6}.
	\end{equation}

	\subsection{Static spacetime: $r_c<\lds$}
	
	For the dS static spacetime, under the coordinate transformation $r=\lds\cos\tilde{\omega}$, $t=\lds\tilde{\tau}$, the metric of the static spacetime can be rewritten as
	\begin{equation}
		\label{staticspacetime}
		ds^2=\lds^2\left(-\sin^2\tilde{\omega}d\tilde{\tau}^2+d\tilde{\omega}^2+\cos^2\tilde{\omega}d\phi^2\right).
	\end{equation}
	For the subregion $A$ defined as $[-\phi_a,\phi_a]$ at $\tilde{\tau}=0$, using the spherical partition function method, one obtains the effective spherical radius as
	\begin{equation}
		r_{\text{eff}}=\lds\cos\tilde{\omega}_c\sin\phi_a<\lds.
	\end{equation}
	By substituting the explicit expression of $r_{\text{eff}}$ into the general expression~\eqref{finite entropy} of the entanglement entropy for the case where the effective spherical radius is smaller than the cosmological horizon, one obtains the entanglement entropy of the subregion $A$:
	\begin{equation}
		S=\frac{c_\ds}{3}\tan^{-1}\left(\frac{\cos\tilde{\omega}_c\sin\phi_a}{\sqrt{1-\cos^2\tilde{\omega}_c\sin^2\phi_a}}\right),
	\end{equation}  
	where $\tilde{\omega}_c$ takes values in the range $(0,\frac{\pi}{2})$. In the case of dS static spacetime, the entanglement entropy of the subregion $A$ is real, which is consistent with the unitarity of the dual field theory. However, the entanglement entropy in this case takes the form of an arctangent function, which, unlike the usual logarithmic divergence of entanglement entropy in local field theories defined at an asymptotic boundary, remains finite. By analyzing the strong subadditivity and the boosted strong subadditivity of entanglement entropy, Refs.~\cite{Miyaji:2015yva,Geng:2019ruz,Geng:2020kxh,Kawamoto:2023nki,Chang:2024voo} showed that the dual field theory exhibits nonlocality in this setup.
		
		For a unitary and local field theory, the entanglement entropy satisfies the strong subadditivity inequality
		\begin{equation}
			\label{bssa}
			S(A) + S(B) - S(A \cup B) - S(A \cap B) \geq 0\,.
		\end{equation}
		By considering infinitesimal boost transformations of subregions $A$ and $B$ within the dS spacetime context, one obtains the differential form of the boosted strong subadditivity~\cite{Chang:2024voo}, which can be written as
		\begin{equation}
			\label{dsdsBSSA}
			S^{\prime\prime}(\Delta\phi)+\frac{1}{\sin\Delta\phi}S^{\prime}(\Delta\phi)\leq 0\,,\quad \Delta\phi=2\phi_a\,,
		\end{equation}
		which is stronger than the standard strong subadditivity constraint.
		
		\begin{figure}[H]
			\centering
			\includegraphics[scale=0.8]{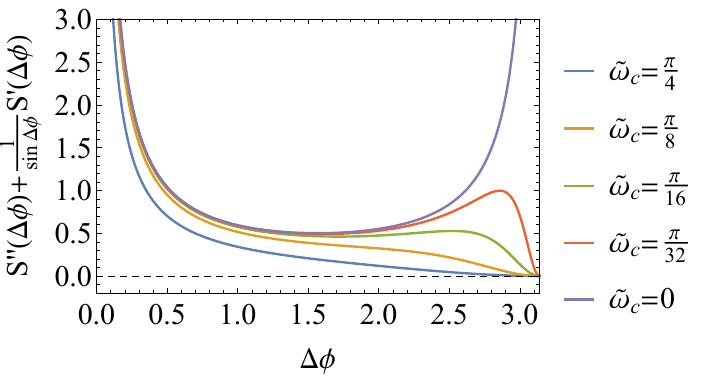}
			\caption{The boosted strong subadditivity. Different colors represent different boundary positions, with the purple line indicating $\tilde{\omega}_c=0$, corresponding to the cosmological horizon $r=\ell_{\mathrm{dS}}$ in dS static spacetime, where the entanglement entropy of the dual field theory strongly violates the boosted strong subadditivity.}
			\label{BSSA}
		\end{figure}
		
		In the static spacetime case, the behavior of the boosted strong subadditivity for different boundary positions is illustrated in Fig.~\ref{BSSA}. It can be observed that in dS static patch holography, the entanglement entropy always violates the boosted strong subadditivity, indicating that the dual field theory defined on the timelike boundary is intrinsically nonlocal. Moreover, as the boundary approaches the cosmological horizon, the violation of the boosted strong subadditivity becomes increasingly pronounced. Finally, we further verify our entanglement entropy results by computing the corresponding geodesic lengths.

	\subsubsection*{Bulk calculation:}
	Similarly, the RT surface area corresponding to the subregion can be computed using embedding coordinates. The relationship between the embedding coordinates and the static spacetime coordinates is given by
	\begin{equation}
		\begin{aligned}
			X&=\lds \cos\tilde{\omega} \cos \phi,\quad  T=\lds\sin \tilde{\omega} \sinh \tilde{\tau},\\
			Y&=\lds \cos\tilde{\omega} \sin \phi,\quad   W=\lds \sin \tilde{\omega} \cosh\tilde{\tau}.
		\end{aligned}
	\end{equation}
	Therefore, the geodesic length connecting the endpoints of the subregion $A$ can be calculated as
	\begin{equation}
		\label{geolengthstatic}
		D_{12}=\lds\cos^{-1}(1-2\cos^2\tilde{\omega}_c\sin^2\phi_a).
	\end{equation}
	Through straightforward trigonometric manipulations and with the relation $c_\ds=\frac{3\lds}{2G}$, one finds that the field theory result matches the gravitational result, $S=\frac{D_{12}}{4G}$.

	\section{Conclusion and Discussion}	
	\label{Conclusion and Discussion}

	In this paper, we employed the Wheeler--DeWitt equation and the Hamiltonian constraint to derive the dual field theory deformations corresponding to the inward motion of the boundary into the bulk in dS spacetime, considering spacelike and timelike boundaries separately. The inward motion of a spacelike boundary corresponds to the $T\bar{T}$ deformation, while that of the timelike boundary corresponds to $T\bar{T}+\Lambda_2$. By analyzing the energy spectrum, we found that for the $T\bar{T}$ deformation, the energy becomes complex when the deformation parameter reaches its critical value $\lambda_0$. By imposing the boundary condition $E_{\lambda_0}^{(T\bar{T})}=E_{\lambda_0}^{(T\bar{T}+\Lambda_2)}$, the $T\bar{T}$ and $T\bar{T}+\Lambda_2$ deformations can be continuously connected, allowing the deformation parameter to continuously approach infinity. The transition from the $T\bar{T}$ to the $T\bar{T}+\Lambda_2$ deformation involves a change of the metric signature from Euclidean to Lorentzian. This coincides with with the transition of the boundary at $r=r_c$ from spacelike to timelike as it crosses the cosmological horizon. This observation motivates our proposal that the composite flow in dS holography is dual to the inward motion of the boundary at $r=r_c$ from the spacelike infinity in the extended spacetime toward the worldline of a static observer.
	
	In the two classes of dS holographic models, the dual field theory is local but non-unitary when it is defined on a spacelike boundary, while it is unitary but nonlocal when it resides on a timelike boundary. The non-unitary and nonlocal properties can be indicated by the behavior of deformed entanglement entropy.   In this paper, we proposed a possible approach toward a unified formulation of dS holography by exploiting the holographic duality of composite flows. The proposal is further supported by the computation of holographic entanglement entropy. By employing the spherical partition function method, we obtained a unified expression \eqref{general entanglement entropy calculate} for the entanglement entropy, which is valid in both  static region $r \in (0,\ell_{\text{ds}})$ and extended spacetime $r \in (\ell_{\text{ds}},\infty)$. Specified to  these two different  regions $r>\ell_{\mathrm{dS}}$ and $r<\ell_{\mathrm{dS}}$, we found two distinct entanglment entropies, corresponding to the pseudo entropy \eqref{psuEE} and the real entropy \eqref{finite entropy}, respectively. These results precisely match the area of the RT surface on the bulk side.

	\subsubsection*{Potential research directions}	
\begin{figure}[H]
		\centering
		\includegraphics[scale=0.67]{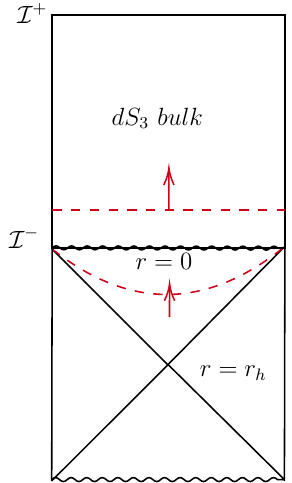}
		\caption{The composite flow formed by the $T\bar{T}+\Lambda_2$ deformation in AdS and the $T\bar{T}$ deformation in dS. Here $r=0$ represents the singularity of the BTZ black hole, which is identified with the past spacelike infinity of the dS spacetime at $\mathcal{I}^{-}$. The radius $r=r_h$ corresponds to the black hole horizon. The red dashed line indicates the spacelike boundary, and the arrows denote the direction of its motion.}
		\label{future1}
	\end{figure}

	It may be possible to help us to  undertand  the properties of black hole singularity by employing the composite flow formed by the two deformation combinations listed in the third column of Table~\ref{summary}. For example, one may construct a model in which the singularity of a non-rotating BTZ black hole is identified with the past infinity of dS spacetime, as illustrated in Fig.~\ref{future1}. In such a setup, the $T\bar{T}+\Lambda_2$ deformation drives the spacelike boundary from the black hole horizon toward the singularity, and at the singularity the flow transitions to the $T\bar{T}$ deformation with a complex deformation parameter, which subsequently pushes the spacelike boundary further into the dS bulk. This suggests an interesting direction for future investigation.
	The full reduction model of dS Jackiw--Teitelboim gravity  contains both the black hole horizon and the cosmological horizon, it is also an interesting direction to investigate the behavior of the composite flow in dS Jackiw--Teitelboim gravity~\cite{Svesko:2022txo,Aguilar-Gutierrez:2024nst}.

	\acknowledgments
	We would like to thank Song He, Shan-Ming Ruan, and Long Zhao for their helpful discussions. This work was supported by the National Natural Science Foundation of China (Grant No.12475056, No.12247101, and No.12105113), the Fundamental Research Funds for the Central Universities (Grant No.lzujbky-2025-jdzx07), the Natural Science Foundation of Gansu Province (No.22JR5RA389, No.25JRRA799), Gansu Province's Top Leading Talent Support Plan, and the `111 Center' under Grant No. B20063. 
	
	

	\bibliographystyle{JHEP}
	
	\bibliography{ref2}
\end{document}